\documentstyle[12pt]{article}
\parskip=\medskipamount                 % space between paragraphs
\thicklines                     % thick straight lines for pictures
\def\Cite#1{\cite{#1}}         % This for Nuclear Physics
\def\Cite#1{{$^{\hbox{\scriptsize \cite{#1}}}$}} % This for Physics Review % %

\def\ordless{{\lower2mm\hbox{$\,\stackrel{\textstyle <}{\sim}\, $}}}
\def\ordgt{{\lower2mm\hbox{$\,\stackrel{\textstyle >}{\sim}\, $}}}

\newcommand{\bim}[6]{\bibitem{#1}#2, {\em #3\/}$\;${\bf #4}$\;$(#5)$\;${#6}.}

\def\startA{\def\theequation{A.\arabic{equation}}}
\def\startB{\def\theequation{B.\arabic{equation}}}
\begin{document}
\edef\@cite#1#2{q{#1\if@tempswa , #2\fi}}
\edef\@biblabel#1{$^#1$\hfill}
\begin{titlepage}
\centerline{\hfill                 NUHEP-TH-91-04}
\vfill
\begin{center}
{\large \bf $R$-matrix Approach to Quantum Superalgebras $su_{q}(m\mid
n)$}\\
\bigskip \centerline{D. Chang$^1$, I. Phillips$^1$, L. Rozansky$^{1,2}$}
\centerline{\em $^{1}$Department of Physics and Astronomy, Northwestern
University,} \centerline{\em Evanston, Illinois 60208}
\centerline{\em $^{2}$Department of Physics, Theory Group, RLM,
University of Texas at Austin,}
\centerline{\em  Austin, Texas 78712}
\vfill
{\bf Abstract}
\end{center}
\begin{quotation}
Quantum superalgebras $su_{q}(m\mid n)$ are studied in the framework
of $R$-matrix formalism. Explicit parametrization of $L^{(+)}$ and
$L^{(-)}$ matrices in terms of $su_{q}(m\mid n)$ generators are
presented. We also show that quantum deformation of nonsimple
superalgebra $su(n\mid n)$ requires its extension to $u(n\mid n)$.

\end{quotation}
\vfill
\end{titlepage}
\pagebreak
\section{Introduction}
\label{1}
In the course of studying quantum algebras, a lot of attention has
been paid to the case of quantum superalgebras (QSA) recently
(see, for example, ref. \cite{13KC}, \cite{KC}).
These algebras provide solutions to the
Yang-Baxter equation and therefore may serve as a source of new
exactly solvable models in statistical mechanics \Cite{S}.
It is also very interesting to study their relation to supergroups
in the WZW models.
Such models were considered in ref.\cite{BO} and their connection with
superconformal models was established.

An intriguing relation between QSA and knot theory was discovered in
ref.\cite{KS}. It was shown there, that the QSA $su_{q}(n\mid n)$ is related
to the Alexander-Conway polynomial in much the same way as the quantum algebra
$su_{q}(n)$ is related to the Jones polynomial. It still remains to be
seen how special properties of the Alexander-Conway polynomial are related
to the nonsimplicity of the superalgebra $su(n\mid n)$.

QSA can also be shown to emerge from the ``new'' solutions to the Yang-Baxter
equation discovered recently in a series of papers in ref.\cite{SB}.
In those papers, these solutions were not recognized as QSA
because of the choice of parametrization of the matrices
$L^{(+)}$ and $L^{(-)}$ in the framework of $R$-matrix formalism.
With proper redefiniton of these parameters, QSA can be easily demonstrated to
be associated with these ``new'' solutions in very much the same way the "old"
solutions are related to the usual quantum algebras.  Here
we develop a convenient parametrization of these matrices for the QSA
$su_{q}(m\mid n)$. In contrast to the previous papers on this subject
\Cite{S,KC}, we use ordinary- (instead of ``super-'') $R$-matrices,
which is in line with the approach of ref.\cite{SB}. Our treatment includes
the case of nonsimple QSA $su_{q}(n\mid n)$, the special
features of which will be displayed.

In following three sections we discuss the QSA $su_{q}(2\mid 0)$,
$su_{q}(0\mid 2)$ and $su_{q}(1\mid 1)$, which are the building blocks of
our general construction. In section V we assemble these blocks in the
matrices $L^{(+)}$ and $L^{(-)}$ for the QSA $su_{q}(m\mid n)$. In
Appendix A we discuss possible choices of the $R$-matrix for that
algebra,
and in Appendix B we give a brief description of the QSA $su_q(2\mid1)$
as a specific example of our general result.

\section{Quantum Superalgebra $su_{q}(2\mid 0)$}
\label{2}
Superalgebra $su_{q}(2\mid 0)$ is, of course, the same as the algebra
$su_{q}(2)$,
which has been described, e.g. in ref.\cite{FRT}. We repeat their analysis
to establish notations that will make it easier to use
as a building block for $su_{q}(m\mid n)$.

According to the $R$-matrix method, one introduces the
upper- and lower diagonal $2\times 2$
matrices $L^{(+)}$ and $L^{(-)}$. Their off-diagonal elements are
raising and lowering generators of $su_{q}(2\mid 0)$ while their
diagonal elements are exponents of Cartan subalgebra generators.

The basic commutation relations between the elements of $L^{(+)}$ and
$L^{(-)}$ are expressed through the following relations\Cite{FRT}:
\begin{equation}
L^{(\pm)}_{2}L^{(\pm)}_{1} =
R_{21}L^{(\pm)}_{1}L^{(\pm)}_{2}R^{-1}_{21},
L_{2}^{(-)}L_{1}^{(+)} = R_{21}L_{1}^{(+)}L_{2}^{(-)}R_{21}^{-1}
\label{2.1}
\end{equation}
Here $L_{1}^{(\pm)}=L^{(\pm)}\otimes 1$, $L_{2}^{(\pm)}=1\otimes
L^{(\pm)}$ and $R_{21}=PRP$, $P$ is a permutation operator, so that
\begin{equation}
R_{21}=PRP=\left(
\begin{array}{cccc}
q& & & \\
&1&q-q^{-1}& \\
&0&1& \\
&&&q
\end{array}
\right)
\label{2.2}
\end{equation}
Eq. (\ref{2.1}) implies the following commutation relations between the
matrix elements of matrices $L^{(\pm)}$:
\begin{equation}
L_{11}^{(\pm)}L_{12}^{(+)}=q^{\mp 1}L_{12}^{(+)}L_{11}^{(\pm)},\;\;
L_{22}^{(\pm)}L_{12}^{(+)}=q^{\pm 1}L_{12}^{(+)}L_{22}^{(\pm)}
\label{2.3}
\end{equation}
\begin{equation}
L_{11}^{(\pm)}L_{21}^{(-)}=q^{\pm 1}L_{21}^{(-)}L_{11}^{(\pm)},\;\;
L_{22}^{(\pm)}L_{21}^{(-)}=q^{\mp 1}L_{21}^{(-)}L_{22}^{(\pm)}
\label{2.4}
\end{equation}
\begin{equation}
\left[L_{12}^{(+)},L_{21}^{(-)}\right]=(q-q^{-1})
\left(L_{11}^{(+)}L_{22}^{(-)}-L_{22}^{(+)}L_{11}^{(-)}\right)
\label{2.5}
\end{equation}
Eqs. (\ref{2.3}) and (\ref{2.4}) show that if $L_{12}^{(+)}$ and
$L_{21}^{(-)}$ are proportional to the raising and lowering operators
$X^{+}$ and $X^{-}$, then both $L_{11}^{(+)}$ and $L_{22}^{(-)}$
should be proportional to $q^{-H}$, while $L_{22}^{(+)}$ and
$L_{11}^{(-)}$ should be proportional to $q^{H}$, where $H$ is a
Cartan subalgebra element. Operators $X^{+}$, $X^{-}$ and $H$ satisfy the
standard commutation relations of $su_{q}(2)$:
\begin{equation}
[H,X^{+}]=X^{+},\;\;\; [H,X^{-}]=-X^{-}
\label{2.6}
\end{equation}
\begin{equation}
[X^{+},X^{-}]=\frac{q^{2H}-q^{-2H}}{q-q^{-1}}
\label{2.7}
\end{equation}
Our normalization for diagonal elements of matrices $L^{(+)}$ and
$L^{(-)}$ differs from that of ref.\cite{FRT}:
\begin{equation}
L_{11}^{(\pm)}=q^{\mp \frac{1}{2}}q^{\mp H},\;\;\;
L_{22}^{(\pm)}=q^{\mp \frac{1}{2}}q^{\pm H}
\label{2.8}
\end{equation}
The advantage of such normalization is that these matrices have simple
forms in the fundamental representation of $su_{q}(2\mid 0)$,
\begin{equation}
L_{11}^{(\pm)}=q^{\mp \left( \begin{array}{cc}
1&0\\
0&0
\end{array} \right)},\;\;\;
L_{22}^{(\pm)}=q^{\mp \left( \begin{array}{cc}
0&0\\
0&1
\end{array} \right)},
\label{2.9}
\end{equation}
and can be easily generalized to the case of $su_{q}(m\mid n)$.

To reconcile eqs. (\ref{2.5}) and (\ref{2.7}), we have to introduce the
factors $(q-q^{-1})$ for $L_{12}^{(+)}$ and $L_{21}^{(-)}$ as well as
an extra negative sign which we ascribe to $L_{12}^{(+)}$ for
reasons that we will explain in Section 5.

Thus we arrive at the following parametrization of the matrices $L^{(+)}$
and $L^{(-)}$:
\begin{equation}
L^{(+)}=\left(
\begin{array}{cc}
q^{-
\left(
\begin{array}{cc}
1&0\\
0&0
\end{array}
\right)}
&(q^{-1}-q)X^{+}\\
0&
q^{-
\left(
\begin{array}{cc}
0&0\\
0&1
\end{array}
\right)}
\end{array}
\right)
\label{2.10}
\end{equation}
\begin{equation}
L^{(-)}=\left(
\begin{array}{cc}
q^{
\left(
\begin{array}{cc}
1&0\\
0&0
\end{array}
\right)}
&0\\
(q-q^{-1})X^{-}&
q^{
\left(
\begin{array}{cc}
0&0\\
0&1
\end{array}
\right)}
\end{array}
\right)
\label{2.11}
\end{equation}
where the matrices of the diagonal blocks should be interpreted in the sense
of eq. (\ref{2.8}) for arbitrary representations.
\section{Quantum Algebra $su_{q}(0\mid 2)$}
\label{3}
The algebra $su_{q}(0\mid 2)$ is, of course, isomorphic to
$su_{q}(2\mid 0)$. However the $R$-matrix that we shall use for this QSA is
different from that of eq. (\ref{2.2}). Actually there are two
choices for this matrix in the literature:
\begin{equation}
R_{21}=\left(
\begin{array}{cccc}
-q^{-1}& & & \\
&\pm 1&q-q^{-1}& \\
&0&\pm 1& \\
&&&-q^{-1}
\end{array}
\right)
\label{3.1}
\end{equation}
The upper signs are advocated in ref.\cite{SB}, while the lower ones -
in ref.\cite{KC}.  We will discuss the relation between these two
possibilities in Appendix A. Here we choose the lower signs, because, as
it will be clear in Section 5, they simplify parametrization of
matrices $L^{(+)}$ and $L^{(-)}$ in terms of QSA generators for the general
case.

With the choice of lower signs in eq.(\ref{3.1}), commutation
relations between matrix elements of $L^{(+)}$ and $L^{(-)}$, which
follow from eq.(\ref{2.1}), are:
\begin{equation}
L_{11}^{(\pm)}L_{12}^{(+)}=q^{\pm 1}L_{12}^{(+)}L_{11}^{(\pm)},\;\;
L_{22}^{(\pm)}L_{12}^{(+)}=q^{\mp 1}L_{12}^{(+)}L_{22}^{(\pm)}
\label{3.2}
\end{equation}
\begin{equation}
L_{11}^{(\pm)}L_{21}^{(-)}=q^{\mp 1}L_{21}^{(-)}L_{11}^{(\pm)},\;\;
L_{22}^{(\pm)}L_{21}^{(-)}=q^{\pm 1}L_{21}^{(-)}L_{22}^{(\pm)}
\label{3.3}
\end{equation}
\begin{equation}
\left[L_{12}^{(+)},L_{21}^{(-)}\right]=(q^{-1}-q)
\left(L_{11}^{(+)}L_{22}^{(-)}-L_{22}^{(+)}L_{11}^{(-)}\right)
\label{3.4}
\end{equation}
As expected this is nothing but eqs.(\ref{2.3}, \ref{2.4}, \ref{2.5})
with $q$ and $q^{-1}$ interchanged.
This time eq. (\ref{3.2}) and (\ref{3.3}) show that if $L_{12}^{(+)}$
and $L_{21}^{(-)}$ are proportional to the raising and lowering operators
$Y^{+}$ and $Y^{-}$, then both $L_{11}^{(+)}$ and $L_{22}^{(-)}$
should be proportional to $q^{J}$, while $L_{11}^{(-)}$ and
$L_{22}^{(+)}$ should be proportional to $q^{-J}$, where $J$ is a
Cartan subalgebra element. As in the case of $su_{q}(2\mid 0)$,
operators $Y^{+}$, $Y^{-}$ and $J$ satisfy commutation relations of
$su_{q}(2)$:
\begin{equation}
[J,Y^{+}]=Y^{+},\;\;\; [J,Y^{-}]=-Y^{-} \label{3.5}
\end{equation}
\begin{equation} [Y^{+},Y^{-}]=\frac{q^{2J}-q^{-2J}}{q-q^{-1}}
\label{3.6}
\end{equation}
Convenient normalization for diagonal elements of $L^{+}$ and
$L^{-}$ is
\begin{equation}
L_{11}^{(\pm)}=q^{\pm \frac{1}{2}}q^{\pm J}=q^{\pm \left(
\begin{array}{cc}
1&0\\
0&0
\end{array}
\right)},\;\;
L_{22}^{(\pm)}=q^{\pm \frac{1}{2}}q^{\mp J}=q^{\pm \left(
\begin{array}{cc}
0&0\\
0&1
\end{array}
\right)},\;\;
\end{equation}
where the $2 \times 2$ matrices are in the fundamental representation.

We multiply all matrix elements of $L^{(+)}$ and $L^{(-)}$ by a factor
$(-1)^{F}$ for future convenience. Here $F$ is a fermionic number
operator with eigenvalues $\frac{1}{2}$ and $-\frac{1}{2}$
respectively in the fundamental representations of $su_{q}(2\mid 0)$
and $su_{q}(0\mid 2)$. Obviously, an extra factor of $(-1)^{F}$ does
not affect the commutation relations (\ref{3.2}-\ref{3.6}), because it
commutes with all operators involved. Thus we get the following
parametrization of $L^{(+)}$ and $L^{(-)}$:
\begin{equation}
L^{(+)}=\left(
\begin{array}{cc}
(-1)^{F}q^{
\left(
\begin{array}{cc}
1&0\\
0&0
\end{array}
\right)}
&(q-q^{-1})(-1)^{F}Y^{+}\\
0&
(-1)^{F}q^{
\left(
\begin{array}{cc}
0&0\\
0&1
\end{array}
\right)}
\end{array}
\right)
\label{3.8}
\end{equation}
\begin{equation}
L^{(-)}=\left(
\begin{array}{cc}
(-1)^{F}q^{-
\left(
\begin{array}{cc}
1&0\\
0&0
\end{array}
\right)}
&0\\
(q^{-1}-q)(-1)^{F}Y^{-}&
(-1)^{F}q^{-
\left(
\begin{array}{cc}
0&0\\
0&1
\end{array}
\right)}
\end{array}
\right)
\label{3.9}
\end{equation}
where the diagonal blocks should be similarly interpreted as functions of the
Cartan subalgebra as in eqs.(\ref{2.10}, \ref{2.11}).
For reasons that will become clear in Section 5, we give an
extra negative sign to $L_{21}^{(-)}$
\section{Quantum Algebra $su_{q}(1\mid 1)$}
\label{4}
We begin, as usual, by presenting the matrix $R_{21}$ for this
algebra:
\begin{equation}
R_{21}=\left(
\begin{array}{cccc}
q& & & \\
&1&q-q^{-1}& \\
&0&1& \\
&&&-q^{-1}
\end{array}
\right)
\label{4.1}
\end{equation}
The corresponding commutation relations for the matrix elements of
$L^{(+)}$ and $L^{(-)}$ are
\begin{equation}
(L_{12}^{(+)})^2=(L_{21}^{(-)})^2=0
\end{equation}
\begin{equation}
L_{11}^{(\pm)}L_{12}^{(+)}=q^{\mp 1}L_{12}^{(+)}L_{11}^{(\pm)},\;\;
L_{22}^{(\pm)}L_{12}^{(+)}=-q^{\mp 1}L_{12}^{(+)}L_{22}^{(\pm)}
\label{4.2}
\end{equation}
\begin{equation}
L_{11}^{(\pm)}L_{21}^{(-)}=q^{\pm 1}L_{21}^{(-)}L_{11}^{(\pm)},\;\;
L_{22}^{(\pm)}L_{21}^{(-)}=-q^{\pm 1}L_{21}^{(-)}L_{22}^{(\pm)}
\label{4.3}
\end{equation}
\begin{equation}
\left[L_{12}^{(+)},L_{21}^{(-)}\right]=(q-q^{-1})
\left(L_{11}^{(+)}L_{22}^{(-)}-L_{22}^{(+)}L_{11}^{(-)}\right)
\label{4.4}
\end{equation}

Algebra $su_{q}(1\mid 1)$ includes three generators $Z^{+}$, $Z^{-}$
and $E$ with (anti-)commutation relations
\begin{equation}
[E,Z^{+}]=[E,Z^{-}]=0
\label{4.5}
\end{equation}
\begin{equation}
\{Z^{+},Z^{-}\}=\frac{q^{2E}-q^{-2E}}{q-q^{-1}}
\label{4.51}
\end{equation}
In the fundamental representation $E=\frac{I}{2}$, where $I$ is identity
operator.

Since the operator $E$ commutes with all generators of $su_{q}(1\mid 1)$,
it is clear that this set of operators is not enough to satisfy eqs.
(\ref{4.2}) and (\ref{4.3}). This deficit is a reflection of the degeneracy
of the Killing scalar product in the superalgebra $su(1\mid 1)$:
\begin{equation}
StrEZ^{+}=StrEZ^{-}=StrE^{2}=0
\label{4.6}
\end{equation}

In order to resolve these difficulties, we introduce another generator which we
shall identify with the fermion number
operator F. It has eigenvalues $\frac{1}{2}$ and $-\frac{1}{2}$ in the
fundamental representation of $su_{q}(1\mid 1)$. The operator $F$ pairs
with $E$ in the Killing scalar product:
\begin{equation}
Str\,EF=\frac{1}{2}
\label{4.7}
\end{equation}
and thus removes the degeneracy.  It
has the following commutation relations with other generators:
\begin{equation}
[F,Z^{+}]=Z^{+},\;\;[F,Z^{-}]=-Z^{-},\;\;[F,E]=0
\label{4.71}
\end{equation}
If we now choose $L_{12}^{(+)}$ and $L_{21}^{(-)}$ to be proportional
respectively to $Z^{+}$ and $Z^{-}$, then we should set
\begin{equation}
L_{11}^{(\pm)}=q^{\mp (F+E)}=q^{\mp \left(
\begin{array}{cc}
1&0\\
0&0
\end{array}
\right)},\;\;
L_{22}^{(\pm)}=(-1)^{F}q^{\mp (F-E)}=(-1)^{F}q^{\pm \left(
\begin{array}{cc}
0&0\\
0&1
\end{array}
\right)},\;\;
\end{equation}
The factor $(-1)^{F}$ in the second expression is responsible for
negative sign in the second formulas of eqs. (\ref{4.2}) and (\ref{4.3}).
A factor of $(-1)^{F}$ added to the expression for $L_{12}^{(+)}$ will
turn the commutator (\ref{4.4}) into the anticommutator (\ref{4.51}). Thus a
complete parametrization of $L^{(+)}$ and $L^{(-)}$ is
\begin{equation}
L^{(+)}=\left(
\begin{array}{cc}
q^{-
\left(
\begin{array}{cc}
1&0\\
0&0
\end{array}
\right)}
&(q-q^{-1})(-1)^{F}Z^{+}\\
0&
(-1)^{F}q^{
\left(
\begin{array}{cc}
0&0\\
0&1
\end{array}
\right)}
\end{array}
\right)
\label{4.9}
\end{equation}
\begin{equation}
L^{(-)}=\left(
\begin{array}{cc}
q^{
\left(
\begin{array}{cc}
1&0\\
0&0
\end{array}
\right)}
&0\\
(q^{-1}-q)Z^{-}&
(-1)^{F}q^{-
\left(
\begin{array}{cc}
0&0\\
0&1
\end{array}
\right)}
\end{array}
\right)
\label{4.10}
\end{equation}
Inclusion of an additional generator $F$ means that we actually
produce the quantum deformation of nonspecial superalgebra $u_{q}(1\mid
1)$ rather than $su_{q}(1\mid 1)$. The same happens to other
nonsimple superalgebras $su_{q}(n\mid n)$ which we discuss in the next
section.

In ref. \cite{liao}, anticommutation relations appear as a result of a graded
tensor product, whereas in refs. \cite{S,KC}, the anticommutation
relations appear through the use of ``super-'' R matrices.  Here, the
ordinary-R matrices and the ordinary
tensor product is used, but the parametrization of the $L^{\pm}$ matrices is
supplemented by extra factors of $(-1)^F$.

\section{Quantum Superalgebra $su_{q}(m\mid n)$}
\label{5}
The possible $R$-matrices for QSA $su_{q}(m\mid n)$ are discussed in
Appendix A. Our choice of matrix $R_{21}$ is a combination of matrices
(\ref{2.2}), (\ref{3.1}) and (\ref{4.1}). This means that if we want
to permute two vectors in the fundamental representation, then we simply
use one of these matrices depending on whether both vectors are
bosonic, fermionic, or one is bosonic and the other is fermionic. The
corresponding $R$-matrix in Eq. (\ref{2.2}) is
\begin{equation}
R=\sum_{I} (-1)^{p_{I}}q^{1-2p_{I}}e_{II}\otimes e_{II}+
\sum_{I\neq J} (-1)^{p_{I}p_{J}}e_{II}\otimes e_{JJ}+
(q-q^{-1})\sum_{I>J} e_{IJ}\otimes e_{JI}
\label{5.1}
\end{equation}
In our notations $e_{IJ}$ is an $(m+n)\times (m+n)$ matrix with only
the $(I,J)$ matrix element being equal to 1, all other matrix elements
are zero.  We also set $p_{I}=0$ for bosons and $p_{I}=1$ for
fermions. We will use indices $I,J,\ldots$ for all $(m+n)$ components of the
fundamental representation, indices $i,j,\ldots$ only for the $m$ bosonic
variables and indices $\alpha ,\beta, \ldots$ only for the $n$ fermionic ones.

We compose matrices $L^{(+)}$ and $L^{(-)}$ out of the matrices
(\ref{2.10}), (\ref{2.11}), (\ref{3.8}), (\ref{3.9}), (\ref{4.9}) and
(\ref{4.10}) in the same way as we composed matrix $R_{21}$:
\begin{eqnarray}
L_{ii}^{(+)}=q^{-e_{ii}},\;\;\;L_{\alpha\alpha}^{(+)}=
(-1)^{F}q^{e_{\alpha\alpha}},
\nonumber\\
L_{i\alpha }^{(+)}=(q-q^{-1})(-1)^{F}Z_{i\alpha }^{+},\;\;\;
L_{ij}^{(+)}=(q^{-1}-q)X_{ij}^{+},\;\;\;i<j\label{5.2}\\
L_{\alpha \beta}^{(+)}=(q-q^{-1})(-1)^{F}Y_{\alpha\beta}^{+}\;\;\;
\alpha<\beta\nonumber\\
L_{IJ}^{(+)}=0,\;\;\;I>J\nonumber
\end{eqnarray}
\begin{eqnarray}
L_{ii}^{(-)}=q^{e_{ii}},\;\;\;L_{\alpha\alpha}^{(-)}=
(-1)^{F}q^{-e_{\alpha\alpha}},
\nonumber\\
L_{\alpha i}^{(-)}=(q^{-1}-q)Z_{\alpha i}^{-},\;\;\;
L_{ij}^{(-)}=(q-q^{-1})X_{ij}^{-},\;\;\;i>j\label{5.3}\\
L_{\alpha \beta}^{(-)}=(q^{-1}-q)(-1)^{F}Y_{\alpha\beta}^{-}\;\;\;
\alpha>\beta\nonumber\\
L_{IJ}^{(-)}=0,\;\;\;I>J\nonumber
\end{eqnarray}

The signs in front of the raising and the lowering operators $X^{\pm}$,
$Y^{\pm}$ and $Z^{\pm}$ are chosen in such a way that when
$q\rightarrow 1$, these operators tend to their classical
counterparts. This means that in the fundamental representation
\begin{equation}
(X,Y,Z)_{IJ}^{\pm}\rightarrow (X_{cl},Y_{cl},Z_{cl})_{IJ}^{\pm}=
e_{IJ}
\label{5.4}
\end{equation}
The generators of the classical superalgebra $su(m\mid n)$ satisfy the
(super-)commutation relations
\begin{equation}
[(X_{cl},Y_{cl},Z_{cl})_{IJ}^{\pm},(X_{cl},Y_{cl},Z_{cl})_{JK}^{\pm}]=
(X_{cl},Y_{cl},Z_{cl})_{IK}^{\pm}
\label{5.5}
\end{equation}
The corresponding relations for the matrix elements of $L^{(+)}$ and
$L^{(-)}$ are
\begin{equation}
\begin{array}{lll}
L_{IJ}^{(+)}L_{JK}^{(+)}&-&(-1)^{P_J(P_I+P_K)}L_{JK}^{(+)}L_{IJ}^{(+)}=\\
                    & &(q^{-1}-q)(-1)^{P_IP_J}L_{JJ}^{(+)}L_{IK}^{(+)}, (I<J<K)
\end{array}
\label{5.6}
\end{equation}
\begin{equation}
\begin{array}{lll}
L_{IJ}^{(-)}L_{JK}^{(-)}&-&(-1)^{P_J(P_I+P_K)}L_{JK}^{(-)}L_{IJ}^{(-)}=\\
                    & &(q-q^{-1})(-1)^{P_KP_J}L_{JJ}^{(-)}L_{IK}^{(-)}, (I>J>K)
\end{array}
\end{equation}
\begin{equation}
\begin{array}{lll}
L_{KI}^{(-)}L_{IJ}^{(+)}&-&(-1)^{P_I(P_J+P_K)}L_{IJ}^{(+)}L_{KI}^{(-)}=\\
& &(q^{-1}-q)(-1)^{P_IP_K}L_{II}^{(+)}L_{KJ}^{(-)}, (I<J<K)
\end{array}
\end{equation}
\begin{equation}
\begin{array}{lll}
L_{KI}^{(-)}L_{IJ}^{(+)}&-&(-1)^{P_I(P_J+P_K)}L_{IJ}^{(+)}L_{KI}^{(-)}=\\
& &(q-q^{-1})(-1)^{P_IP_J}L_{KJ}^{(+)}L_{II}^{(-)}, (I<K<J)
\end{array}
\end{equation}
\begin{equation}
\begin{array}{ll}
L_{KJ}^{(-)}L_{IK}^{(+)}-&(-1)^{P_K(P_I+P_J)}L_{IK}^{(+)}L_{KJ}^{(-)}=\\
&\left\{ \begin{array}{ll}
(q^{-1}-q)(-1)^{P_IP_K}L_{IJ}^{(+)}L_{KK}^{(-)} & (I<J<K)\\
(q-q^{-1})(-1)^{P_JP_K}L_{KK}^{(+)}L_{IJ}^{(-)} & (J<I<K)
 \end{array} \right.
\end{array}
\end{equation}
\begin{equation}
L_{JI}^{(-)}L_{JK}^{(+)}=
(-1)^{P_KP_I}(-1)^{P_J}q^{1-2P_J}L_{JK}^{(+)}L_{JI}^{(-)},
  (I<J<K)
\end{equation}
\begin{equation}
L_{KJ}^{(-)}L_{IJ}^{(+)}=
(-1)^{P_IP_K}(-1)^{P_J}q^{-1+2P_J}L_{IJ}^{(+)}L_{KJ}^{(-)},
  (I<J<K)
\label{5.7}
\end{equation}
Matching the classical limit of the eqs. (\ref{5.6}-\ref{5.7}) with the
classical
commutator (\ref{5.5}) dictates the choice between factors
$(q-q^{-1})$ and $(q^{-1}-q)$ for the operators $X^{\pm}$ and $Y^{\pm}$.
The signs of the operators $Z^{\pm}$ are not prescribed by these
requirements and can be chosen arbitrarily.

The exponents of $q$ appearing in diagonal elements
$L_{II}^{(+)}$ and $L_{JJ}^{(-)}$ are not supertraceless. Therefore,
strictly speaking, they are not the elements of $su_{q}(m\mid n)$ Cartan
subalgebra. To overcome this difficulty we multiply these matrix
elements by factors of $q^{\frac{1}{n-m}}$ and $q^{\frac{1}{m-n}}$
respectively. Such factors will render the exponents supertraceless
without affecting the commutation relations (\ref{2.3}-\ref{2.5}),
(\ref{3.2}-\ref{3.4}) and (\ref{4.2}-\ref{4.4}).

This ends the process of parametrization of matrix elements of
$L^{(+)}$ and $L^{(-)}$ in terms of
$su_{q}(m\mid n)$ generators if $m \neq n$.
However supertracelesness can not be
achieved if $m=n$. In this case the condition of supertracelesness of the
original classical algebra $su(n\mid n)$ should be dropped, so that we
deal in fact with algebra $u(n\mid n)$. Its Cartan subalgebra includes
one generator with nonvanishing supertrace which can be identified with the
fermion
number operator $F$. If $m\neq n$, F can be considered to be just an
element of $su_{q}(m\mid n)$ Cartan subalgebra.

\section{Conclusion}
\label{6}

We considered the construction of quantum superalgebra $su_{q}(m\mid n)$
in the framework of $R$-matrix formalism. In contrast to the papers
\cite{KC} and \cite{S}, we used ordinary (not super-) commutation
relations between matrix elements of $L^{(+)}$ and $L^{(-)}$ while
parametrizing them in terms of the generators of QSA $su_{q}(m\mid n)$.
Thus it can easily be shown that the ``special'' solutions of
Yang-Baxter equation,
discussed in ref.\cite{SB}, are related to QSA in the same way as
ordinary solutions are related to quantum algebras through the Reshetikhin
construction \cite{R}. Therefore we conjecture that QSA can be used to generate
all possible solutions to the Yang-Baxter equation.

Our study of nonsimple superalgebras $su(n\mid n)$ also revealed that their
quantum deformation requires extending them to superalgebras $u(n\mid
n)$, whose Cartan subalgebras include the fermionic number operator.

\subsection*{Acknowledgement}

It is pleasure for us to acknowledge the stimulating conversations with Profs.
L. Kauffman and H. Saleur on quantum superalgebras and Alexander polynomial.
D.C. also wishes to thank the Institute of Physics at Academia Sinica for
its hospitality while this manuscript was finalized.
This work was supported in part by the U.S. Department of Energy.

\subsection*{Appendix A -- Choice of $R$-matrix for QSA $su_{q}(m\mid n)$}
\setcounter{equation}{0}
\startA

Here we discuss possible choices of $R$-matrix for QSA $su_{q}(m\mid
n)$. To simplify discussion, we will consider $\check{R}$-matrix,
which is the product of permutation operator $P$ and original
$R$-matrix:
\begin{equation}
\check{R}=PR
\label{A.1}
\end{equation}
The $\check{R}$-matrix, presented in ref.\cite{KC}, is
\begin{equation}
\check{R}=\sum_{I\neq J}(-1)^{p_{I}p_{J}}e_{JI}\otimes e_{IJ}+
\sum_{I}(-1)^{p_{I}}q^{1-2p_{I}}e_{II}\otimes e_{II}+
(q-q^{-1})\sum_{I<J}e_{II}\otimes e_{JJ}
\label{A.2}
\end{equation}
However, the $\check{R}$-matrix, in ref.\cite{SB}
for ``nonstandard''
solutions of the Yang-Baxter equation, has a slightly different form:
\begin{equation}
\check{R}^{\prime}=\sum_{I\neq J}e_{JI}\otimes e_{IJ}+
\sum_{I}(-1)^{p_{I}}q^{1-2p_{I}}e_{II}\otimes e_{II}+
(q-q^{-1})\sum_{I<J}e_{II}\otimes e_{JJ}
\label{A.3}
\end{equation}
The factor of $(-1)^{p_{I}p_{J}}$ in the first sum of formula
(\ref{A.2}) seems to be quite reasonable: an extra negative sign
arises through permutation of two fermionic vectors. This is why the
$R_{21}$-matrix in eq.(\ref{3.1}) with lower signs, which stems from the
$\check{R}$-matrix in eq.(\ref{A.2}), provides simple parametrization of
$L^{(+)}$ and $L^{(-)}$ (the upper signs in eq.(\ref{3.1}) would correspond to
the $\check{R}$-matrix (\ref{A.3})).

To show the equivalence of the $\check{R}$-matrices in eqs.(\ref{A.2}) and
(\ref{A.3}), let us consider the action of $\check{R}$-matrix on the
tensor product $\bigotimes_{k=1}^{N}V_{k}$ of $N$ fundamental
representations $V_{k}$ (N is any integer). Let us denote by
$e_{I_{k}}$ ($1\leq I_{k}\leq m+n$) the basis vectors of $V_{k}$.

Consider now operator $D$ which calculates the parity of number of
``fermionic disorders'' in basis vectors of tensor product:
\begin{equation}
D\bigotimes_{k=1}^{N}e_{I_{k}}=\bigotimes_{k=1}^{N}(-1)^
{d_{\{I_{k}\}}}e_{I_{k}}
\label{A.4}
\end{equation}
Here $d_{\{I_{k}\}}$ is the number of fermionic disorders, i.e.
$d_{\{I_{k}\}}$ counts the number of pairs of indices $k,l$, such that
\begin{equation}
k<l,\;\;\;m,I_{l},I_{k}
\label{A.5}
\end{equation}
Obviously, D commutes with operators $e_{II}\otimes e_{JJ}$ for all
possible values of $I$ and $J$, because these operators do not
permute different vectors. $D$ commutes also with operators
$e_{jI}\otimes e_{Ij}$, because they permute pairs of vectors at
least one of which is bosonic. However
\begin{equation}
De_{\alpha\beta}\otimes e_{\beta\alpha}D^{-1}=
-e_{\alpha\beta}\otimes e_{\beta\alpha},\;\;\; \alpha\neq\beta
\label{A.6}
\end{equation}
because operator $e_{\alpha\beta}\otimes e_{\beta\alpha}$ permutes two
different fermionic vectors, thus changing the parity of fermionic
disorder number. Therefore we see that
\begin{equation}
D\sum_{I\neq J}(-1)^{p_{I}p_{J}}e_{JI}\otimes e_{IJ}D^{-1}=
\sum_{I\neq J}e_{JI}\otimes e_{IJ}
\label{A.7}
\end{equation}
and
\begin{equation}
D\check{R}D^{-1}=\check{R}^{\prime}
\label{A.8}
\end{equation}
Eq.(\ref{A.8}) shows equivalence of two $\check{R}$-matrices
(\ref{A.2}) and (\ref{A.3}).

\subsection*{Appendix B -- Quantum Superalgebra $su_q(2\mid1)$}
\setcounter{equation}{0}
\startB

As an example we give a brief description of the $su_q(2\mid1)$ quantum
superalgebra.  The matrices $L^{\pm}$ are parametrized as in eqs.
(\ref{5.2}),(\ref{5.3}),

\begin{equation}
L^+=
q^{-1}
\left(
\begin{array}{ccc}
q^{h_2}  &  (q^{-1}-q)\alpha^+  &  (q-q^{-1})(-1)^Fb^+\\
0        &  q^{h_1+h_2} &  (q-q^{-1})(-1)^F\beta^+\\
0        &  0           & (-1)^Fq^{h_1+2h_2}
\end{array}
\right)
\end{equation}

\begin{equation}
L^-=
q
\left(
\begin{array}{ccc}
q^{-h_2}  &  0           &  0 \\
(q-q^{-1})\alpha^- & q^{-h_1-h_2} &  0 \\
(q^{-1}-q)b^-     & (q^{-1}-q)\beta^-    & (-1)^Fq^{-h_1-2h_2}
\end{array}
\right)
\end{equation}

Eqs. (\ref{2.1}) and (\ref{5.1}) imply
\begin{equation}
\begin{array}{lr}
 b^+\alpha^+-q\alpha^+b^+=0 & {[\beta^+,\alpha^+]}=-b^+q^{h_1+h_2} \\
 {[\beta^-,\alpha^-]}=b^-q^{-(h_1+h_2)} & \beta^+b^++q^{-1}b^+\beta^+=0 \\
 \beta^-b^-+q^{-1}b^-\beta^-=0 & \alpha^+b^--b^-\alpha^+=-\beta^-q^{h_2} \\
 \alpha^+\beta^--q\beta^-\alpha^+=0 & \alpha^-b^+-b^+\alpha^-=\beta^+q^{-h_2}
\\
 b^+\beta^-+\beta^-b^+=
\alpha^+q^{-h_1-2h_2} & \beta^+\alpha^--q^{-1}\alpha^-\beta^+=0\\
 \beta^+b^-+b^-\beta^+=\alpha^-q^{h_1+2h_2}.&
\end{array}
\end{equation}

These relations reduce to the classical relations of $su(2\mid1)$ in
the classical limit.  A different parametrization can be found in Ref.
\cite{KC}

\section*{Addendum}

After this paper was submitted, H. Saleur brought to our attention the
paper ref. \cite{liao} which addresses the problem of $su_{q}(1\mid 1)$.
The authors of ref. \cite{liao} use a graded tensor product in order to
produce anticommutators in some of the relations of (\ref{2.1}). We
use an ordinary tensor product, which is in line with the approach taken in
ref. \cite{SB}. However, we supplement their parametrization of
$L^{\pm}$ matrices with an extra factor of $(-1)^{F}$ to achieve the
same effect.


\begin{thebibliography}{99}
\bim{13KC}{P. Kulish, N. Reshetikhin}{Lett.Math.Phys.}{18}{1989}{143}
\bim{KC}{M. Chaichian, P. Kulish}{Phys.Lett.}{234B}{1990}{72}
\bim{S}{S. Saleur}{Nucl.Phys.}{B336}{1990}{363}
\bim{BO}{M. Bershadsky, H. Ooguri}{Phys.Lett.}{229B}{1989}{374}
\bibitem{KS} L. Kauffman, H. Saleur, {\em Free Fermions and the
Alexander-Conway Polynomial\/}, preprint EFI 90-42.
\bim{SB}{N. Jing, M.-L. Ge, Y.-S. Wu}{Lett.Math.Phys.}{21}{1991}{193}\\
Y. Cheng, M.-L. Ge, K. Xue, {\em New Solutions of Yang-Baxter
Equation\/}, preprint ITP-SB-90-38 (1990);
\bibitem{FRT} L. Faddeev, N. Reshetikhin, L. Takhtajan, {\em
Quantization of Lie Groups and Lie Algebras\/}, preprint LOMI E-14-87
(1987)
\bim{liao}{Li-Liao, Xing-Chang Song}{Mod.Phys.Lett.}{A6}{1991}{959}
\bibitem{R} N. Reshetikhin, {\em Quantized Universal Enveloping Algebras, the
Yang-Baxter Equation and Invariants of Links}, parts I and II, preprint
LOMI E-4-87, E-17-87.
\end{thebibliography}
\end{document}